\documentclass[amsmath,aps,showpacs,a4paper,10pt]{revtex4}

 \usepackage{epsf}
 \usepackage{graphicx}    
 \usepackage{amssymb}     

\begin{document}

 \newcommand{\be}[1]{\begin{equation}\label{#1}}
 \newcommand{\ee}{\end{equation}}
 \newcommand{\bea}{\begin{eqnarray}}
 \newcommand{\eea}{\end{eqnarray}}
 \def\disp{\displaystyle}

 \def\gsim{ \lower .75ex \hbox{$\sim$} \llap{\raise .27ex \hbox{$>$}} }
 \def\lsim{ \lower .75ex \hbox{$\sim$} \llap{\raise .27ex \hbox{$<$}} }

 \begin{titlepage}

 \begin{flushright}
 arXiv:1511.00376
 \end{flushright}

 \title{\Large \bf Exact Cosmological Solutions of $f(R)$
 Theories via~Hojman~Symmetry}

 \author{Hao~Wei\,}
 \email[\,email address:\ ]{haowei@bit.edu.cn}
 \affiliation{School of Physics, Beijing Institute
 of Technology, Beijing 100081, China}

 \author{Hong-Yu~Li\,}
 \affiliation{School of Physics, Beijing Institute
 of Technology, Beijing 100081, China}

 \author{Xiao-Bo~Zou}
 \affiliation{School of Physics, Beijing Institute
 of Technology, Beijing 100081, China}

 \begin{abstract}\vspace{1cm}
 \centerline{\bf ABSTRACT}\vspace{2mm}
 Nowadays, $f(R)$ theory has been one of the leading modified
 gravity theories to explain the current accelerated expansion
 of the universe, without invoking dark energy. It is of
 interest to find the exact cosmological solutions of $f(R)$
 theories. Besides other methods, symmetry has been proved as
 a powerful tool to find exact solutions. On the other hand,
 symmetry might hint the deep physical structure of a theory,
 and hence considering symmetry is also well motivated. As is
 well known, Noether symmetry has been extensively used in
 physics. Recently, the so-called Hojman symmetry was also
 considered in the literature. Hojman symmetry directly deals
 with the equations of motion, rather than Lagrangian or
 Hamiltonian, unlike Noether symmetry. In this work, we
 consider Hojman symmetry in $f(R)$ theories in both the metric
 and Palatini formalisms, and find the corresponding exact
 cosmological solutions of $f(R)$ theories via Hojman symmetry.
 There exist some new solutions significantly different from
 the ones obtained by using Noether symmetry in $f(R)$
 theories. To our knowledge, they also have not been found
 previously in the literature. This work confirms that Hojman
 symmetry can bring new features to cosmology and gravity theories.
 \end{abstract}

 \pacs{98.80.-k, 04.50.Kd, 11.30.-j, 95.36.+x}

 \maketitle

 \end{titlepage}

 \renewcommand{\baselinestretch}{1.0}


\section{Introduction}\label{sec1}

The current accelerated expansion of the universe could be due
 to an unknown energy  component (dark energy) or a modification to
 general relativity (modified gravity)~\cite{r1,r2}. In the
 literature, various modified gravity theories were proposed
 to account for the cosmic acceleration, such as $f(R)$
 theory~\cite{r2,r3,r4,r45,r48}, scalar-tensor theory~\cite{r4,r5},
 braneworld model~\cite{r6,r7}, Galileon gravity~\cite{r8,r9},
 Gauss-Bonnet gravity~\cite{r10}, $f(T)$ theory~\cite{r11,r12},
 massive gravity~\cite{r13,r14}. Nowadays, modified gravity
 theories have been one of the main fields in modern cosmology.

As one of the leading modified gravity theories, $f(R)$ theory
 was proposed by generalizing the well-known Einstein-Hilbert
 Lagrangian $R$ used in general relativity (GR) to an arbitrary
 function $f(R)$, where $R$ is the scalar curvature. In fact,
 $f(R)$ theory has been extensively studied in the literature
 for many years (see e.g.~\cite{r2,r3,r4} for reviews). It can
 be used to drive inflation (see e.g.~\cite{r15}), play the
 role of dark matter (see e.g.~\cite{r16}), or drive the
 current accelerated expansion of the universe as
 an competitive alternative of dark energy (see
 e.g.~\cite{r17,r18}).

Note that there exist two different types of $f(R)$ theories
 in the literature (see e.g.~\cite{r2,r3,r4}), namely $f(R)$
 theory in the metric formalism, and $f(R)$ theory in the
 Palatini formalism. In the metric formalism, the affine
 connection $\Gamma^\lambda_{\alpha\beta}$ depends on the
 metric $g_{\mu\nu}$, and hence the field equations are
 derived by the variation of the action with respect to the
 metric $g_{\mu\nu}$ only. On the other hand, in the Palatini
 formalism, the affine connection
 $\Gamma^\lambda_{\alpha\beta}$ and the metric $g_{\mu\nu}$
 are treated as independent variables when one varies the
 action. As is well known, in the case of GR
 (namely $f(R)\propto R$), the field equations are completely
 identical in these two formalisms. However, in the case
 of non-linear $f(R)$, the field equations are different
 in these two formalisms. So, the metric and Palatini $f(R)$
 theories should be considered separately.

It is of interest to find the exact cosmological solutions of
 $f(R)$ theories. Besides other methods (e.g.
 reconstruction~\cite{r47}), symmetry has been proved as a
 powerful tool to find exact solutions. On the other hand,
 symmetry might hint the deep physical structure of a theory,
 and hence considering symmetry is also well motivated. As is
 well known, Noether symmetry has been extensively used in
 cosmology and gravity theories, for instance, scalar field
 cosmology~\cite{r19,r20}, $f(R)$ theory~\cite{r21,r22,r23,
 r24,r25,r46,r48}, scalar-tensor theory~\cite{r26,r27}, $f(T)$
 theory~\cite{r28,r29}, Gauss-Bonnet gravity~\cite{r30},
 non-minimally coupled cosmology~\cite{r31}, and others~\cite{r32}.
 It is worth noting that a (point-like) Lagrangian should
 be given {\it a priori} when one uses Noether symmetry.

In this work, we are interested to consider the so-called
 Hojman symmetry in $f(R)$ theories, and find the corresponding
 exact cosmological solutions of $f(R)$ theories via Hojman
 symmetry. Unlike Noether conservation theorem, the symmetry
 vectors and the corresponding conserved quantities in Hojman
 conservation theorem can be obtained by using the equations
 of motion directly, without using Lagrangian or Hamiltonian.
 In general, its conserved quantities and the exact solutions
 can be quite different from the ones via Noether symmetry.
 In fact, recently Hojman symmetry has been used in cosmology
 and gravity theory~\cite{r33,r34,r35}. It is found that
 Hojman symmetry exists for a wide range of the potential
 $V(\phi)$ of quintessence~\cite{r33} and scalar-tensor
 theory~\cite{r34}, and the corresponding exact cosmological
 solutions have been obtained. While Noether symmetry exists
 only for exponential potential $V(\phi)$~\cite{r19,r26,r27},
 Hojman symmetry can exist for a wide range of potentials
 $V(\phi)$, including not only exponential but also power-law,
 hyperbolic, logarithmic and other complicated
 potentials~\cite{r33,r34}. On the other hand, it is also
 found that Hojman symmetry exists in $f(T)$ theory and the
 corresponding exact cosmological solutions are
 obtained~\cite{r35}. The functional form of $f(T)$ is
 restricted to be the power-law or hypergeometric type, while
 the universe experiences a power-law or hyperbolic expansion.
 These results are also different from the ones obtained by
 using Noether symmetry in $f(T)$ theory~\cite{r28}. Therefore,
 although some exact cosmological solutions of $f(R)$ theories
 were found by using Noether symmetry in the
 literature~\cite{r21,r22,r23,r24,r25}, it is still interesting
 to find them by using Hojman symmetry instead, because as
 mentioned above one can expect that the solutions via Hojman
 symmetry might be quite different. On the other hand, as an
 important lesson from history, we consider that it is better
 to keep an open mind to this new symmetry, especially if
 it can bring something new.

The so-called Hojman symmetry was proposed in the
 year 1992~\cite{r36}. Following~\cite{r36} and
 e.g.~\cite{r33,r34,r35}, we consider a set of second order
 differential equations
 \be{eq1}
 \ddot{q}^{\, i}={\cal F}^i\left(q^j,\,\dot{q}^j,\,t\right),~~~~~~~
 i, j=1, \ldots, n,
 \ee
 where a dot denotes a derivative with respect to time $t$.
 If $X^i=X^i\left(q^j,\,\dot{q}^j,\,t\right)$ is a symmetry
 vector for Eq.~(\ref{eq1}), it satisfies~\cite{r37,r38}
 \be{eq2}
 \frac{d^2X^i}{dt^2}-
 \frac{\partial {\cal F}^i}{\partial q^j}X^j
 -\frac{\partial {\cal F}^i}{\partial\dot{q}^j}\frac{dX^j}{dt}=0\,,
 \ee
 where
 \be{eq3}
 \frac{d}{dt}=\frac{\partial}{\partial t}
 +\dot{q}^i\frac{\partial}{\partial q^i}+
 {\cal F}^i\frac{\partial}{\partial\dot{q}^i}\,.
 \ee
 The symmetry vector $X^i$ is defined so that the infinitesimal
 transformation
 \be{eq4}
 \hat{q}^{\,i}=q^i+\epsilon X^i\left(q^j,\,\dot{q}^j,\,t\right)
 \ee
 maps solutions $q^i$ of Eq.~(\ref{eq1}) into solutions
 $\hat{q}^{\,i}$ of the same equations (up to $\epsilon^2$
 terms)~\cite{r37,r38}. If the ``force'' ${\cal F}^i$ satisfies
 (in some coordinate systems)
 \be{eq5}
 \frac{\partial {\cal F}^i}{\partial\dot{q}^i}
 = -\frac{d}{dt}\ln\gamma\,,
 \ee
 where $\gamma=\gamma(q^i)$ is a function of $q^i$, then
 \be{eq6}
 Q=\frac{1}{\gamma}\frac{\partial\left(\gamma X^i\right)}
 {\partial q^i}+\frac{\partial}{\partial\dot{q}^i}
 \left(\frac{dX^i}{dt}\right)
 \ee
 is a conserved quantity for Eq.~(\ref{eq1}), namely $dQ/dt=0$.
 Note that in the case of $\gamma=const.$, Eqs.~(\ref{eq5}) and
 (\ref{eq6}) become simple and trivial. In the proof of Hojman
 conservation theorem~\cite{r36} (see also e.g.~\cite{r39}),
 neither a Lagrangian nor a Hamiltonian is needed, and no
 previous knowledge of a constant of motion for
 system~(\ref{eq1}) is invoked either~\cite{r36}. In this
 way, Hojman conservation theorem is different from Noether
 conservation theorem.

In the present work, we consider Hojman symmetry in $f(R)$
 theories in both the metric and Palatini formalisms, and
 find the corresponding exact cosmological solutions. In fact,
 they are the main contents of Secs.~\ref{sec2} and \ref{sec3},
 respectively. One can expect new results by using Hojman
 symmetry in $f(R)$ theories. The brief concluding remarks
 are given in Sec.~\ref{sec4}.


\section{Exact cosmological solutions of $f(R)$ theory
 in~the~metric~formalism}\label{sec2}

In this section, we consider $f(R)$ theory in the metric
 formalism at first. The action is given by
 \be{eq7}
 {\cal S}=\frac{1}{2\kappa^2}\int d^4 x\sqrt{-g}\,f(R)+{\cal S}_M\,,
 \ee
 where $\kappa^2\equiv 8\pi G$, $g$ is the  determinant of the
 metric $g_{\mu\nu}$, and ${\cal S}_M$ is the matter action.
 In the metric formalism, the affine connection
 $\Gamma^\lambda_{\alpha\beta}$ depends on the metric $g_{\mu\nu}$,
 and hence the field equations are derived by the variation of
 the action with respect to the metric $g_{\mu\nu}$. Throughout
 this work, we consider a spatially flat Friedmann-Robertson-Walker
 (FRW) universe whose spacetime is described by
 \be{eq8}
 ds^2=-dt^2+a^2(t)\,d{\bf x}^2\,,
 \ee
 where $a$ is the scale factor. For this metric, in the metric
 formalism, the Ricci scalar $R$ is given by~\cite{r2,r3,r4}
 \be{eq9}
 R=6\left(2H^2+\dot{H}\right)\,,
 \ee
 where $H\equiv\dot{a}/a$ is the Hubble parameter, and a dot
 denotes a derivative with respect to cosmic time $t$. The
 modified Friedmann equations read~\cite{r2,r3,r4}
 \bea
 &&3FH^2=\left(FR-f\right)/2-3H\dot{F}+\kappa^2 \rho_{_M}
 \,,\label{eq10}\\[2mm]
 &&-2F\dot{H}=\ddot{F}-H\dot{F}+\kappa^2\left(\rho_{_M}
 +p_{_M}\right)\,,\label{eq11}
 \eea
 where $F=f_{,R}\equiv\partial f/\partial R$, and $\rho_{_M}$,
 $p_{_M}$ are the energy density and pressure of matter,
 respectively. The energy conservation equation of matter is
 given by
 \be{eq12}
 \dot{\rho}_{_M}+3H\left(\rho_{_M}+p_{_M}\right)=0\,.
 \ee
 The equation-of-state parameter (EoS) of matter is defined by
 $w_{_M}=p_{_M}/\rho_{_M}\,$. In particular, $w_{_M}=0$
 and $1/3$ correspond to pressureless matter
 and radiation, respectively.

The main difficulty to consider Hojman symmetry in the metric
 $f(R)$ theory is that the corresponding equations of motion
 (Eqs.~(\ref{eq10}) and (\ref{eq11})) are 4th order with
 respect to the scale factor $a$, while Hojman symmetry deals
 with 2nd order equations as mentioned in Sec.~\ref{sec1}. We
 should try to recast them as second order differential
 equations. Inspired by the well-known conformal
 transformation~\cite{r2,r3,r4}, we introduce new variables
 $\tilde{t}$ and $\tilde{a}$ according to
 \be{eq13}
 d\tilde{t}=\sqrt{F}\,dt\,,~~~~~~~\tilde{a}=\sqrt{F}\,a\,.
 \ee
 While the traditional conformal transformation mainly deals
 with the Lagrangian/action, here we instead directly deal with the
 equations of motion using Eq.~(\ref{eq13}). Also, we introduce
 \be{eq14}
 \tilde{H}\equiv\frac{1}{\tilde{a}}
 \frac{d\tilde{a}}{d\tilde{t}}=\frac{1}{\sqrt{F}}\left(H+
 \frac{\dot{F}}{2F}\right)\,,
 \ee
 in which we have used Eq.~(\ref{eq13}). Introducing a new
 scalar field $\phi$ according to
 \be{eq15}
 \kappa\phi = \sqrt{\frac{3}{2}}\,\ln F\,,
 \ee
 we can recast Eq.~(\ref{eq10}) as
 \be{eq16}
 \tilde{H}^2=\frac{\kappa^2}{3}\left(\tilde{\rho}_\phi+
 \tilde{\rho}_{_M}\right)\,,
 \ee
 where $\tilde{\rho}_{_M}=\rho_{_M}/F^2$, and
 \be{eq17}
 \tilde{\rho}_\phi=\frac{1}{2}\left(\frac{d\phi}{d\tilde{t}}
 \right)^2+V(\phi)\,,~~~~~~~~
 V(\phi)=\frac{FR-f}{2\kappa^2F^2}\,.
 \ee
 By the help of Eqs.~(\ref{eq9}) and (\ref{eq10}), we can
 recast Eq.~(\ref{eq11}) as the equation of motion for the
 scalar field $\phi$, namely
 \be{eq18}
 \frac{d^2\phi}{d\tilde{t}^2}+
 3\tilde{H}\frac{d\phi}{d\tilde{t}}
 +V_{,\phi}=\frac{\kappa}{\sqrt{6}}\left(
 \tilde{\rho}_{_M}-3\tilde{p}_{_M}\right)\,,
 \ee
 where $\tilde{p}_{_M}=p_{_M}/F^2$,
 and $V_{,\phi}=\partial V /\partial\phi\,$.
 In fact, Eq.~(\ref{eq18}) is equivalent to
 \be{eq19}
 \frac{d\tilde{\rho}_\phi}{d\tilde{t}}+3\tilde{H}
 \left(\tilde{\rho}_\phi+\tilde{p}_\phi\right)=
 \frac{\kappa}{\sqrt{6}}\left(\tilde{\rho}_{_M}-3\tilde{p}_{_M}
 \right)\frac{d\phi}{d\tilde{t}}\,,
 \ee
 where
 \be{eq20}
 \tilde{p}_\phi=\frac{1}{2}\left(\frac{d\phi}{d\tilde{t}}
 \right)^2-V(\phi)\,.
 \ee
 On the other hand, Eq.~(\ref{eq12}) can be recast as
 \be{eq21}
 \frac{d\tilde{\rho}_{_M}}{d\tilde{t}}+3\tilde{H}\left(
 \tilde{\rho}_{_M}+\tilde{p}_{_M}\right)=
 -\frac{\kappa}{\sqrt{6}}\left(
 \tilde{\rho}_{_M}-3\tilde{p}_{_M}\right)\frac{d\phi}{d\tilde{t}}\,.
 \ee
 So, the ``total energy conservation equation'' holds, namely
 \be{eq22}
 \frac{d\tilde{\rho}_{tot}}{d\tilde{t}}+3\tilde{H}\left(
 \tilde{\rho}_{tot}+\tilde{p}_{tot}\right)=0\,,
 \ee
 where $\tilde{\rho}_{tot}=\tilde{\rho}_\phi+\tilde{\rho}_{_M}$, and
 $\tilde{p}_{tot}=\tilde{p}_\phi+\tilde{p}_{_M}$. Using
 Eqs.~(\ref{eq16}) and (\ref{eq22}), we obtain
 \be{eq23}
 \frac{1}{\tilde{a}}\frac{d^2\tilde{a}}{d\tilde{t}^2}=-
 \frac{\kappa^2}{6}\left(\tilde{\rho}_{tot}
 +3\tilde{p}_{tot}\right)\,.
 \ee

Now, we try to consider Hojman symmetry in the metric $f(R)$
 theory. Following~\cite{r33,r34,r35}, we introduce a new
 variable $\tilde{x}\equiv\ln\tilde{a}$. From now on, in order
 to make the expressions simple, we use an empty circle
 ``$\circ$'' to denote a derivative with respect to the
 ``new time'' $\tilde{t}$. So, it is easy to see
 $\mathring{\tilde{x}}=\tilde{H}$. Using Eqs.~(\ref{eq23}) and
 (\ref{eq16}), we have
 \be{eq24}
 \stackrel{\,_{\circ\circ}}{\tilde{x}}\ =\mathring{\tilde{H}}
 =\frac{d^2\tilde{x}}{d\tilde{t}^2}
 =-\frac{\kappa^2}{2}\left(\tilde{\rho}_\phi+\tilde{p}_\phi
 +\tilde{\rho}_{_M}+\tilde{p}_{_M}\right)\,.
 \ee
 Following~\cite{r33,r34}, here we only consider the ``dark
 energy'' dominated epoch, and hence the contributions from
 matter can be ignored. For convenience, we also set the
 unit $\kappa=1$. Noting Eqs.~(\ref{eq17}) and (\ref{eq20}),
 it is easy to see that Eq.~(\ref{eq24}) becomes
 \be{eq25}
 \stackrel{\,_{\circ\circ}}{\tilde{x}}\ =-s(\tilde{x})\,
 \mathring{\tilde{x}}^2={\cal F}(\tilde{x},\mathring{\tilde{x}})\,,
 \ee
 where
 \be{eq26}
 s(\tilde{x})=\frac{1}{2}\phi^{\prime\,2}(\tilde{x})\,,
 \ee
 and a prime denotes a derivative with respect to the variable
 of the function, namely $h^\prime(y)=dh(y)/dy$. Note that
 Eqs.~(\ref{eq25}) and (\ref{eq26}) in this work have the same
 forms as Eqs.~(21) and (22) of~\cite{r33}, except for the
 different variables $\tilde{x}$ and $\tilde{t}$. Therefore,
 the derivations below are straightforward by using the needed
 results from~\cite{r33}. If Hojman symmetry exists in this
 theory, the condition~(\ref{eq5}) should be satisfied.
 From Eqs.~(\ref{eq25}), (\ref{eq5})
 and (\ref{eq3}) replaced $t$ with $\tilde{t}$, we find that
 \be{eq27}
 \gamma(\tilde{x})=\gamma_0\exp
 \left(2\int s(\tilde{x})\,d\tilde{x}\right)\,,
 \ee
 where $\gamma_0$ is an integration constant.
 Following~\cite{r33,r34,r35}, we assume that the symmetry
 vector $X$ does not explicitly depend on time $\tilde{t}$.
 Then, Eq.~(\ref{eq2}) replaced $t$
 with $\tilde{t}$ becomes~\cite{r33}
 \be{eq28}
 \left[s(\tilde{x})\frac{\partial X}{\partial\tilde{x}}+
 s^\prime (\tilde{x})X+\frac{\partial^2 X}
 {\partial\tilde{x}^2}\right]+\mathring{\tilde{x}}^2 s^2 (\tilde{x})
 \frac{\partial^2 X}{\partial \mathring{\tilde{x}}^2} -
 \mathring{\tilde{x}}\left[2s(\tilde{x})\frac{\partial^2 X}
 {\partial\tilde{x}\partial\mathring{\tilde{x}}}+
 s^\prime (\tilde{x})\frac{\partial X}{\partial
 \mathring{\tilde{x}}}\right]=0\,.
 \ee
 Using Eqs.~(\ref{eq16}), (\ref{eq18}), and ignoring the
 contributions from matter, we have~\cite{r33}
 \be{eq29}
 \frac{V^\prime(\phi)}{V(\phi)}=\frac{s(\tilde{x})\phi^\prime
 (\tilde{x})-\phi^{\prime\prime}(\tilde{x})-3\phi^\prime
 (\tilde{x})}{3-\frac{1}{2}\phi^{\prime\,2}(\tilde{x})}\,,
 \ee
 which is useful to derive the potential $V(\phi)$.


\subsection{Power-law solution}\label{sec2a}

In fact, the differential equation for the symmetry vector $X$,
 namely Eq.~(\ref{eq28}), is difficult to solve in general. The
 authors of~\cite{r33} had tried various ansatz
 for the symmetry vector $X(\tilde{x},\mathring{\tilde{x}})$.
 For the ansatz
 \be{eq30}
 X(\tilde{x},\mathring{\tilde{x}})=
 \mathring{\tilde{x}}\,g(\tilde{x})\,,~~~~~~~{\rm and}~~~~~~~
 g(\tilde{x})=\lambda\exp\left(\frac{\alpha^2}{2}\tilde{x}\right)\,,
 \ee
 the corresponding cosmological solutions
 obtained in~\cite{r33} are given by
 \bea
 &&V(\varphi)=\frac{2\left(6-\alpha^2\right)Q_0^2}{\alpha^4
 \lambda^2}\,e^{\mp\alpha\varphi}\,,\label{eq31}\\
 &&\tilde{a}(\tilde{t})=e^{\tilde{x}_0}\left[1+
 \frac{Q_0 \alpha^2}{2\lambda}\exp\left(-\frac{\alpha^2}{2}
 \tilde{x}_0\right)\left(\tilde{t}-\tilde{t}_0
 \right)\right]^{2/\alpha^2}\,,\label{eq32}\\
 &&\varphi(\tilde{t})=\pm\frac{2}{\alpha}\ln\left[1+\frac{Q_0
 \alpha^2}{2\lambda}\exp\left(-\frac{\alpha^2}{2}
 \tilde{x}_0\right)\left(\tilde{t}-\tilde{t}_0\right)\right]
 \,,\label{eq33}
 \eea
 where $\varphi=\phi-\phi_0$, and $\phi_0$, $\tilde{t}_0$,
 $\tilde{x}_0$, $\lambda$, $\alpha$ are constants. The
 conserved quantity is given by~\cite{r33}
 \be{eq34}
 \mathring{\tilde{x}}g^\prime(\tilde{x})=Q_0=const.
 \ee
 We refer to~\cite{r33} for the detailed derivations. With
 these results, we can convert them into the cosmological
 solutions in the metric $f(R)$ theory. For convenience, we
 recast Eq.~(\ref{eq31}) as $V(\phi)=V_0\,e^{\mp\alpha\phi}$, where
 $V_0=2(6-\alpha^2)Q_0^2\exp(\pm\phi_0)/(\alpha^4\lambda^2)$
 is constant. Using this $V(\phi)$ and Eqs.~(\ref{eq17}),
 (\ref{eq15}), we have
 \be{eq35}
 FR-f=2V_0 F^\beta\,,
 \ee
 where $\beta=2\mp\sqrt{\frac{3}{2}}\,\alpha$. Noting that
 $F=f_{,R}\equiv\partial f/\partial R$, it is a differential
 equation for $f(R)$ with respect to $R$ in fact. It is easy
 to find the solution as
 \be{eq36}
 f(R)=c_1 R^n\,,
 \ee
 where $n=\beta/(\beta-1)$ and
 $c_1=((n-1)/(2V_0 n^\beta))^{1/(\beta-1)}$ are both constants.
 In the case of $n=1$, the solution reads
 $f(R)=c_2 R-2c_2^\beta V_0$ where $c_2$ is constant. Since it
 is trivial, we do not consider the case of $n=1$ any more. Let
 us turn to find the scale factor $a(t)$ and the Hubble
 parameter $H(t)$ or $H(a)$. Using Eqs.~(\ref{eq15}) and
 (\ref{eq33}), we get
 \be{eq37}
 \sqrt{F}=\exp\left(\frac{\phi_0}{\sqrt{6}}
 \right)\left[1+c_0\left(\tilde{t}-\tilde{t}_0
 \right)\right]^{\pm\sqrt{\frac{2}{3}}/\alpha}\,,
 \ee
 where $c_0=(Q_0
 \alpha^2/(2\lambda))\exp(-\alpha^2 \tilde{x}_0 /2)$ is
 constant. Integrating $dt=d\tilde{t}/\sqrt{F}$
 from Eq.~(\ref{eq13}) gives
 \be{eq38}
 t-t_0=c_{31}\left[1+c_0\left(\tilde{t}-\tilde{t}_0\right)
 \right]^{1\mp\sqrt{\frac{2}{3}}/\alpha}\,,~~~{\rm or}
 ~~~1+c_0\left(\tilde{t}-\tilde{t}_0\right)=c_{32}
 \left(t-t_0\right)^{1/(1\mp\sqrt{\frac{2}{3}}/\alpha)}\,,
 \ee
 where $c_{31}=\exp(-\phi_0/\sqrt{6})/(c_0(1\mp\sqrt{2/3}/\alpha))$,
 $c_{32}=c_{31}^{-1/(1\mp\sqrt{2/3}/\alpha)}$, and $t_0$ is an
 integration constant. Substituting Eq.~(\ref{eq38}) into
 Eqs.~(\ref{eq32}), (\ref{eq37}) and
 then $a=\tilde{a}/\sqrt{F}$ from Eq.~(\ref{eq13}), we obtain
 \be{eq39}
 a(t)=c_3\left(t-t_0\right)^m\,,
 \ee
 where
 $m=(2/\alpha^2\mp\sqrt{2/3}/\alpha)/(1\mp\sqrt{2/3}/\alpha)$
 and $c_3=\exp(\tilde{x}_0-\phi_0/\sqrt{6})
 \,c_{32}^{2/\alpha^2\mp\sqrt{2/3}/\alpha}$ are both constants.
 Obviously, the universe  experiences a power-law expansion.
 Note that this solution can also be found via Noether
 symmetry~\cite{r21,r22,r23}. From Eq.~(\ref{eq39}), it is easy
 to obtain the Hubble parameter as
 \be{eq40}
 H(t)=\frac{\dot{a}}{a}=m\left(t-t_0\right)^{-1}\,,~~~~~~~
 {\rm or}~~~~~~~H(a)=H_0\,a^{-1/m}\,,
 \ee
 where $H_0=mc_3^{1/m}$ is the Hubble constant.


\subsection{New solutions}\label{sec2b}

In~\cite{r33}, other ansatz for the symmetry vector $X$ are
 also considered. For the ansatz
 \be{eq41}
 X=X(\mathring{\tilde{x}})=A_0\,\mathring{\tilde{x}}^{-1/\alpha}\,,
 \ee
 the corresponding cosmological solutions
 obtained in~\cite{r33} are given by
 \bea
 &&V(\varphi)=\lambda\varphi^{-4\alpha}-\frac{8}{3}\lambda
 \alpha^2\varphi^{-4\alpha-2}\,,\label{eq42}\\[0.7mm]
 &&\tilde{a}(\tau)=e^{\alpha s_0}\exp\left(\alpha\left((1+
 \alpha)\tau\right)^{1/(1+\alpha)}\right)\,,\label{eq43}\\[1mm]
 &&\varphi(\tau)=\mp\sqrt{8}\,\alpha
 \left[(1+\alpha)\tau\right]^{1/(2(1+\alpha))}\,,\label{eq44}
 \eea
 where $\varphi=\phi-\phi_c$,
 $\tau=y_0+\alpha^{-1}\left|Q_0\right|^{-\alpha}\tilde{t}$,
 and $\phi_c$, $y_0$, $s_0$, $\lambda$, $\alpha$ are constants.
 The conserved quantity is given by~\cite{r33}
 \be{eq45}
 \frac{\mathring{\tilde{x}}^{-1/\alpha}}{s_0
 -\tilde{x}/\alpha}=Q_0=const.
 \ee
 Note that the same solutions
 (\ref{eq42})---(\ref{eq44}) can also be found by
 using another ansatz~\cite{r33}
 \be{eq46}
 X(\tilde{x},\mathring{\tilde{x}})=
 \mathring{\tilde{x}}\,g(\tilde{x})\,,~~~~~~~{\rm and}~~~~~~~
 g(\tilde{x})=\frac{(f_0+\tilde{x})^{1+\alpha}}{1+\alpha}\,.
 \ee
 We refer to~\cite{r33} for the detailed derivations. With
 these results, we can convert them into the cosmological
 solutions in the metric $f(R)$ theory. Using
 Eqs.~(\ref{eq15}) and (\ref{eq44}), we get
 \be{eq47}
 \sqrt{F}=\exp\left(\frac{\phi_c}{\sqrt{6}}
 \right)\exp\left(c_0\tilde{\tau}^\beta\right)\,,
 \ee
 where $\beta=1/(2(1+\alpha))$, $c_0=\mp 2\alpha/\sqrt{3}$ are
 both constants,
 and $\tilde{\tau}=(1+\alpha)\tau=\tilde{y}_0+\tilde{Q}_0\tilde{t}$.
 Substituting Eq.~(\ref{eq47}) into $d\tilde{t}=\sqrt{F}\,dt$
 from Eq.~(\ref{eq13}), we have
 \be{eq48}
 \frac{d\tilde{\tau}}{dt}=c_{21}\exp\left(
 c_0\tilde{\tau}^\beta\right)\,,
 \ee
 where $c_{21}=\tilde{Q}_0\exp(\phi_c/\sqrt{6})$ is constant.
 The solution of Eq.~(\ref{eq48}) is given by
 \be{eq49}
 t-t_0=-\frac{\tilde{\tau}}{\beta c_{21}}
 E_{\frac{\beta-1}{\beta}}\left(c_0\tilde{\tau}^\beta\right)\,,
 \ee
 where $t_0$ is an integration
 constant, and $E_n(z)=\int_1^\infty e^{-zu\,}u^{-n\,}du$
 is the exponential integral function. Unfortunately, if
 $\beta\not=1$, it is hard to find $\tilde{\tau}$ as a function
 of $t$ by solving Eq.~(\ref{eq49}), and hence it is also hard
 to find the scale factor $a$ as a function of $t$. Therefore,
 we only consider the case of $\beta=1$ here (note that
 $\beta=1$ corresponds to $\alpha=-1/2$ actually). In this
 case, the solution of Eq.~(\ref{eq48}) reads
 \be{eq50}
 \tilde{\tau}=-\frac{1}{c_0}\ln\left(t-t_0\right)+c_2\,,
 \ee
 where $c_2=-c_0^{-1}\ln(-c_0 c_{21})$ is
 constant. Substituting Eq.~(\ref{eq50}) into
 Eqs.~(\ref{eq43}), (\ref{eq47}) and then
 $a=\tilde{a}/\sqrt{F}$ from Eq.~(\ref{eq13}), noting that $\beta=1$
 (namely $\alpha=-1/2$), we have
 \be{eq51}
 a(t)=c_3\left(t-t_0\right)^{1+c_2/c_0}\exp
 \left(-\frac{1}{2c_0^2}\left(\,\ln(t-t_0)\right)^2\right)\,,
 \ee
 where $c_3=\exp(-s_0/2-\phi_c/\sqrt{6}-c_0c_2-c_2^2/2)$ is
 constant. {\em Obviously, it is not a power-law solution. To
 our knowledge, this new solution has not been found previously
 in the literature.} From Eq.~(\ref{eq51}), it is easy to
 obtain the Hubble parameter as
 \be{eq52}
 H(t)=\frac{\dot{a}}{a}=c_0^{-2}\left(t-t_0\right)^{-1}
 \left(c_4-\ln\left(t-t_0\right)\right)\,,
 \ee
 where $c_4=c_0(c_0+c_2)$ is constant. From Eq.~(\ref{eq51}),
 we find
 \be{eq53}
 \ln\left(t-t_0\right)=\eta(a)=c_4\pm\sqrt{c_4^2-2\ln\frac{a}{c_3}}\,.
 \ee
 Substituting it into Eq.~(\ref{eq52}), we get
 \be{eq54}
 H(a)=c_0^{-2} \left(c_4-\eta(a)\right) e^{-\eta(a)}\,.
 \ee
 Note that in the case of $\beta=1$ (namely $\alpha=-1/2$),
 the corresponding $c_0^{-2}=3$ in fact. Next, let us turn to
 find $f(R)$ as a function of $R$. Using Eqs.~(\ref{eq17}),
 (\ref{eq15}), and (\ref{eq42}) with $\alpha=-1/2$, we obtain
 \be{eq55}
 f=FR-2\lambda F^2\left[\left(\sqrt{\frac{3}{2}}
 \,\ln F-\phi_c\right)^2-\frac{2}{3}\right]\,.
 \ee
 Noting that $F=f_{,R}\equiv\partial f/\partial R$, it is a
 differential equation for $f(R)$ with respect to $R$.
 Unfortunately, this differential equation is hard to solve in
 general. Since the solution $f(R)=c_{10}R+c_{20}$ is trivial,
 we should find a way to obtain the non-trivial solution.
 Substituting Eq.~(\ref{eq52}) into Eq.~(\ref{eq9}), we get
 \be{eq56}
 R(t)=-6c_0^{-4}\left(t-t_0\right)^{-2}\left[
 -2c_4^2+c_0^2(1+c_4)-\ln\left(t-t_0\right)\left(c_0^2-
 4c_4+2\ln\left(t-t_0\right)\right)\right]\,.
 \ee
 Substituting Eq.~(\ref{eq50}) into Eq.~(\ref{eq47}) with
 $\beta=1$, we obtain
 \be{eq57}
 \sqrt{F}=\exp\left(\frac{\phi_c}{\sqrt{6}}
 \right)\exp\left(c_0c_2-\ln\left(t-t_0\right)\right)\,.
 \ee
 Substituting Eqs.~(\ref{eq56}) and (\ref{eq57})
 into Eq.~(\ref{eq55}), $f(t)$ can be found as a function of
 time $t$. Unfortunately, it is hard to solve Eq.~(\ref{eq56})
 and obtain $t-t_0$ or $\ln(t-t_0)$ as an explicit function of
 $R$. So, we cannot obtain $f(R)$ as an explicit function of
 $R$. Nevertheless, with $f(t)$ and $R(t)$, we can still regard
 $f(t)=f(t(R))=f(R)$ as an implicit function in principle.

Note that other exotic ansatz for the symmetry vector $X$ are
 considered in~\cite{r33}, and the corresponding $V(\phi)$,
 $\tilde{a}(\tilde{t})$ and $\phi(\tilde{t})$ are found. However, it
 is hard to obtain $f(R)$, $a(t)$ and $H(t)$ in these cases.
 Although further complicated ansatz for the symmetry vector
 $X$ beyond the ones in~\cite{r33} could be tried, we stop here. Let
 us turn to $f({\cal R})$ theory in the Palatini formalism.


\section{Exact cosmological solutions of $f({\cal R})$ theory
 in~the~Palatini~formalism}\label{sec3}

In this section, we consider $f({\cal R})$ theory in the
 Palatini formalism. The action is given by
 \be{eq58}
 {\cal S}=\frac{1}{2\kappa^2}
 \int d^4 x\sqrt{-g}\,f({\cal R})+{\cal S}_M\,,
 \ee
 where $\kappa^2\equiv 8\pi G$, $g$ is the  determinant of the
 metric $g_{\mu\nu}$, and ${\cal S}_M$ is the matter action.
 In the Palatini formalism, the affine connection
 $\Gamma^\lambda_{\alpha\beta}$ and the metric $g_{\mu\nu}$
 are treated as independent variables. So, the Ricci scalar
 $\cal R$ is different from the one in the metric formalism,
 and their relation reads~\cite{r2,r3,r4,r24,r25,r40,r41,r42}
 \be{eq59}
 {\cal R}=R+\frac{3}{2F^2}\nabla_\mu F \nabla^\mu
 F-\frac{3}{F}\Box F\,,          
 \ee
 where $F=f_{,\cal R}\equiv\partial f/\partial {\cal R}$. Thus,
 in the Palatini formalism ${\cal R}\not=R=6(2H^2+\dot{H})$
 generally. We consider a spatially flat FRW universe whose
 spacetime is described by Eq.~(\ref{eq8}), and the modified
 Friedmann equations read~\cite{r2,r3,r4,r24,r25,r40,r41,r42}
 \bea
 F{\cal R}-2f=-\kappa^2\left(\rho_{_M}-3p_{_M}
 \right)\,,\label{eq60}\\[1mm]
 \disp 6F\left(H+\frac{\dot{F}}{2F}\right)^2-f
 =\kappa^2\left(\rho_{_M}+3p_{_M}\right)\,.\label{eq61}
 \eea
 The energy conservation equation of matter is given by
 Eq.~(\ref{eq12}). In this section, we do not ignore the
 contributions from matter. To be simple, here we only consider the
 case of $w_{_M}=p_{_M}/\rho_{_M}=0$, namely pressureless
 matter. Differentiating Eq.~(\ref{eq60}) and
 using Eq.~(\ref{eq12}) with $p_{_M}=0$, one has~\cite{r2,r3,r4,r40}
 \be{eq62}
 \dot{\cal R}=\frac{3\kappa^2 H\rho_{_M}}
 {F_{,\cal R}{\cal R}-F}=-3H\cdot\frac{F{\cal R}
 -2f}{F_{,\cal R}{\cal R}-F}\,.
 \ee

Similar to Sec.~\ref{sec2}, we introduce new variables
 $\tilde{t}$ and $\tilde{a}$ according to Eq.~(\ref{eq13}),
 and obtain $\tilde{H}$ in Eq.~(\ref{eq14}) by definition.
 Adding Eqs.~(\ref{eq60}) and (\ref{eq61}) with $p_{_M}=0$,
 we have
 \be{eq63}
 \tilde{H}^2=\frac{3f-F{\cal R}}{6F^2}\,.
 \ee
 Following~\cite{r33,r34,r35}, we introduce a new
 variable $\tilde{x}\equiv\ln\tilde{a}$, and use an empty
 circle ``$\circ$'' to denote a derivative with respect
 to the ``new time'' $\tilde{t}$. So, it is easy to see
 $\mathring{\tilde{x}}=\tilde{H}$. Since the right hand side
 of Eq.~(\ref{eq63}) is a function explicitly depends only
 on $\cal R$. Thus, ${\cal R}={\cal R}(\mathring{\tilde{x}})$
 is a function explicitly depends only on
 $\mathring{\tilde{x}}=\tilde{H}$.
 Differentiating Eq.~(\ref{eq63}) with respect to $\tilde{t}$,
 and using Eqs.~(\ref{eq62}), (\ref{eq13}), (\ref{eq14}), we obtain
 \be{eq64}
 \mathring{\tilde{H}}=\frac{F{\cal R}-2f}{2F^2}\,.
 \ee
 Noting that ${\cal R}={\cal R}(\mathring{\tilde{x}})$, its
 right hand side is a function explicitly depends only on
 $\mathring{\tilde{x}}$. Therefore, Eq.~(\ref{eq64}) is in fact
 \be{eq65}
 \stackrel{\,_{\circ\circ}}{\tilde{x}}\ =\mathring{\tilde{H}}
 =\frac{F{\cal R}-2f}{2F^2}={\cal F}(\mathring{\tilde{x}})\,.
 \ee
 The ``force'' $\cal F$ explicitly depends only
 on $\mathring{\tilde{x}}=\tilde{H}$. If Hojman symmetry
 exists in the Palatini $f({\cal R})$ theory, the condition
 (\ref{eq5}) should be satisfied. Noting Eq.~(\ref{eq3}) and
 $\gamma=\gamma(\tilde{x})$, we recast Eq.~(\ref{eq5}) as
 \be{eq66}
 -\frac{1}{\mathring{\tilde{x}}}\frac{\partial {\cal F}
 (\mathring{\tilde{x}})}{\partial\mathring{\tilde{x}}}=
 \frac{\partial}{\partial\tilde{x}}\ln\gamma(\tilde{x})\,.
 \ee
 Since its left hand side is a function of
 $\mathring{\tilde{x}}$ only, and its right hand side is a
 function of $\tilde{x}$ only, they must be equal to a same
 constant in order to ensure that Eq.~(\ref{eq66}) always
 holds. For convenience, we let this constant be $2n$, and
 then Eq.~(\ref{eq66}) can be separated into two ordinary
 differential equations
 \be{eq67}
 \frac{\partial}{\partial\tilde{x}}\ln\gamma(\tilde{x})=2n\,,
 ~~~~~~~\frac{\partial {\cal F}(\mathring{\tilde{x}})}{\partial
 \mathring{\tilde{x}}}=-2n\mathring{\tilde{x}}\,.
 \ee
 Thus, it is easy to find that
 \bea
 &&\gamma(\tilde{x})=\gamma_0\,e^{2n\tilde{x}}\,,\label{eq68}\\[1mm]
 &&{\cal F}(\mathring{\tilde{x}})=
 -n\mathring{\tilde{x}}^2+c_0\,,\label{eq69}
 \eea
 where $\gamma_0$ and $c_0$ are both integration constants.
 In the following subsections, we consider the cases of $c_0=0$
 and $c_0\not=0$, respectively.


\subsection{Power-law solution with $c_0=0$}\label{sec3a}

In the case of $c_0=0$, substituting Eq.~(\ref{eq69}) into
 Eq.~(\ref{eq65}), and using Eq.~(\ref{eq63}), we obtain
 \be{eq70}
 3(n-2)f=(n-3)F{\cal R}\,.
 \ee
 Noting $F=f_{,\cal R}\equiv\partial f/\partial {\cal R}$, it
 is a differential equation for $f({\cal R})$ with respect to
 $\cal R$ in fact. Note that if $n=0$, from Eq.~(\ref{eq70}) we
 have $F{\cal R}-2f=0$, while $F{\cal R}-2f=-\kappa^2\rho_{_M}$
 from Eq.~(\ref{eq60}). So, $n\not=0$ is required unless
 $\rho_{_M}=0$. In the case of $n=2$, $f({\cal R})=const.$, and
 in the case of $n=3$, $f({\cal R})=0$. Thus, we do not
 consider these trivial cases of $n=2$ and $3$. In other cases,
 the solution of Eq.~(\ref{eq70}) is given by
 \be{eq71}
 f({\cal R})=c_1{\cal R}^m\,,
 \ee
 where $m=3(n-2)/(n-3)$ and $c_1$ are both
 constants. Substituting Eq.~(\ref{eq69}) into Eq.~(\ref{eq65})
 with $c_0=0$, we have $\mathring{\tilde{H}}=-n\tilde{H}^2$
 whose solution reads
 \be{eq72}
 \tilde{H}(\tilde{t})=\frac{1}{n}\left(\tilde{t}+c_2\right)^{-1}\,,
 \ee
 where $c_2$ is an integration constant. From
 $\tilde{H}=\mathring{\tilde{a}}/\tilde{a}$, it is easy to get
 \be{eq73}
 \tilde{a}(\tilde{t})=c_3\left(\tilde{t}+c_2\right)^{1/n}\,,
 \ee
 where $c_3$ is an integration constant. Substituting
 Eq.~(\ref{eq71}) into Eq.~(\ref{eq63}), and
 using Eq.~(\ref{eq72}), we have
 \be{eq74}
 {\cal R}=\left[\frac{6c_1 m^2}{n^2(3-m)}\right]^{1/(2-m)}
 \left(\tilde{t}+c_2\right)^{-2/(2-m)}\,,
 \ee
 and then
 \be{eq75}
 F=c_1 m{\cal R}^{m-1}=c_{41}\left(\tilde{t}
 +c_2\right)^{-2(m-1)/(2-m)}\,,
 \ee
 where $c_{41}=c_1 m[6c_1 m^2/(n^2(3-m))]^{(m-1)/(2-m)}$ is
 constant. Since $n\not=0$ and $n\not=2$ as mentioned above,
 we note that $m\not=0$, $m\not=2$ and $m\not=3$. Substituting
 Eq.~(\ref{eq75}) into $dt=d\tilde{t}/\sqrt{F}$
 from Eq.~(\ref{eq13}), it is easy to obtain
 \be{eq76}
 t-t_0=\frac{nc_{41}^{-1/2}}{3-n}\left(\tilde{t}+
 c_2\right)^{(3-n)/n}\,,~~~~~~~{\rm or}~~~~~~~
 \tilde{t} + c_2 = c_{42}\left(t-t_0\right)^{n/(3-n)}\,,
 \ee
 where $c_{42}=[c_{41}^{1/2}(3/n-1)]^{n/(3-n)}$ is constant.
 Substituting Eq.~(\ref{eq75}) into $a=\tilde{a}/\sqrt{F}$ from
 Eq.~(\ref{eq13}), and using Eqs.~(\ref{eq73}), (\ref{eq76}),
 we find that
 \be{eq77}
 a(t)=c_4\left(t-t_0\right)^{2m/3}\,,
 \ee
 where $c_4=c_3 c_{41}^{-1/2}c_{42}^{4/n-2}$ is constant.
 Obviously, the universe  experiences a power-law expansion.
 Note that this solution can also be found via Noether
 symmetry~\cite{r24,r25}. From Eq.~(\ref{eq77}), it is easy
 to get the Hubble parameter as
 \be{eq78}
 H(t)=\frac{\dot{a}}{a}=\frac{2m}{3}\left(t-t_0\right)^{-1}\,,
 ~~~~~~~{\rm or}~~~~~~~H(a)=H_0\,a^{-3/(2m)}\,,
 \ee
 where $H_0=(2m/3)\,c_4^{3/(2m)}$ is the Hubble constant.

Let us turn to the conserved quantity. Following~\cite{r33,r34,r35},
 we assume that the symmetry vector $X$ does not explicitly
 depend on time. Substituting Eq.~(\ref{eq69}) with $c_0=0$
 into Eq.~(\ref{eq2}), the equation for $X$ reads
 \be{eq79}
 \frac{\partial^2 X}{\partial\tilde{x}^2}
 -2n\mathring{\tilde{x}}\frac{\partial^2 X}{\partial\tilde{x}
 \partial\mathring{\tilde{x}}}+n^2\mathring{\tilde{x}}^2
 \frac{\partial^2 X}{\partial\mathring{\tilde{x}}^2}+n
 \frac{\partial X}{\partial\tilde{x}}=0\,.
 \ee
 To solve this equation, we adopt the ansatz
 \be{eq80}
 X=A_0\mathring{\tilde{x}}^\alpha e^{\beta\tilde{x}}+A_1\,,
 \ee
 where $A_0$, $A_1$, $\alpha$, $\beta$ are all constants,
 and $\alpha$, $\beta$ cannot be zero at the same time.
 Substituting Eq.~(\ref{eq80}) into Eq.~(\ref{eq79}), we
 find that the solutions have $n\alpha-\beta=0$
 or $n\alpha-\beta=n$. Substituting Eqs.~(\ref{eq80}) and
 (\ref{eq68}) into Eq.~(\ref{eq6}), the conserved quantity
 $Q$ is given by
 \be{eq81}
 Q=2nA_1-(2+\alpha)(n\alpha-\beta-
 n)A_0\mathring{\tilde{x}}^\alpha e^{\beta\tilde{x}}\,.
 \ee
 If $n\alpha-\beta=n$ or $\alpha=-2$, then $Q=2nA_1=const.$
 is trivial. If $n\alpha-\beta=0$ and $\alpha\not=-2$, we get
 \be{eq82}
 \mathring{\tilde{x}}e^{n\tilde{x}}=const.
 \ee
 In fact, this conserved quantity can be found in another way.
 Noting that $\mathring{\tilde{x}}=\tilde{H}$,
 $\tilde{x}=\ln\tilde{a}$, and using Eqs.~(\ref{eq72}),
 (\ref{eq73}), one can find the same conserved quantity
 given in Eq.~(\ref{eq82}) again. This can be regarded as a
 confirmation of Hojman conservation theorem.


\subsection{New solutions with $c_0\not=0$}\label{sec3b}

In the case of $c_0\not=0$, substituting Eq.~(\ref{eq69}) into
 Eq.~(\ref{eq65}), and using Eq.~(\ref{eq63}), we obtain
 \be{eq83}
 3(n-2)f=6c_0 F^2+(n-3)F{\cal R}\,.
 \ee
 Noting $F=f_{,\cal R}\equiv\partial f/\partial {\cal R}$, it
 is a differential equation for $f({\cal R})$ with respect to
 $\cal R$ in fact. If $n=2$, its solutions are $f({\cal R})=const.$
 or $f({\cal R})={\cal R}^2/(12c_0)+const.$, which lead to
 $F{\cal R}-2f=const.$, while $F{\cal R}-2f=-\kappa^2\rho_{_M}$
 from Eq.~(\ref{eq60}). So, $n\not=2$ is required
 unless $\rho_{_M}=const.$ If $n=3$, the solution reads
 $f({\cal R})={\cal R}^2/(8c_0)\pm c_1{\cal R}/\sqrt{8c_0}+c_1^2/4$,
 which is trivial. If $n=3/2$, the solution is given by
 $f({\cal R})=c_1{\cal R}+c_2$, which reduces to GR in fact.
 Besides these dismissed cases, we consider the
 cases of $n\not=0$ and $n=0$ one by one in the followings.
 In fact, some new solutions can be found via Hojman symmetry.


\subsubsection{The case of $n\not=0$}\label{sec3b1}

In the case of $n\not=0$ (and also $n\not=2$, $3$, $3/2$ as
 mentioned above), it is hard to solve Eq.~(\ref{eq83}) and
 obtain $f({\cal R})$ as an explicit function of $\cal R$.
 In fact, from Eq.~(\ref{eq83}), $f({\cal R})$ and $\cal R$
 satisfy the equation
 \bea
 2\left(2-\frac{3}{n}\right){\rm arctanh}\left[
 \frac{(2n-3){\cal R}}{\xi\left(f({\cal R}),{\cal R}\right)}\right]
 &+&\left(2-\frac{3}{n}\right)\ln\left[24c_0 f({\cal R})
 -n{\cal R}^2\right]\nonumber\\[1mm]
 &-&2\left(1-\frac{3}{n}\right)\ln\left[2\left((n-3){\cal R}+
 \xi\left(f({\cal R}),{\cal R}\right)\right)
 \right]=const.\,,\label{eq84}
 \eea
 or another equation
 \be{eq85}
 2\left(\frac{3}{n}-1\right)\ln f({\cal R})-{\rm
 left\ hand\ side\ of\ Eq.~(\ref{eq84})}=const.,
 \ee
 where $\xi\left(f({\cal R}),{\cal R}\right)=
 \left[(n-3)^2{\cal R}^2+72c_0(n-2)f({\cal R})\right]^{1/2}$.
 Using Eqs.~(\ref{eq84}) or (\ref{eq85}), we can regard
 $f({\cal R})$ as an implicit function of
 $\cal R$ in principle. Let us move forward. Substituting
 Eq.~(\ref{eq69}) into Eq.~(\ref{eq65}), we have
 $\mathring{\tilde{H}}=-n\tilde{H}^2+c_0$, whose solution
 for $c_0\not=0$ and $n\not=0$ is given by
 \be{eq86}
 \tilde{H}(\tilde{t})=\sqrt{\frac{c_0}{n}}\,\tanh
 \left(\sqrt{nc_0}\ \tilde{t}+c_2\right)\,,
 \ee
 where $c_2$ is an integration constant. From
 $\tilde{H}=\mathring{\tilde{a}}/\tilde{a}$, it is easy to get
 \be{eq87}
 \tilde{a}(\tilde{t})=c_3\left[\cosh
 \left(\sqrt{nc_0}\ \tilde{t}+c_2\right)\right]^{1/|n|}\,,
 \ee
 where $c_3$ is an integration constant. On the other hand, it
 is hard to solve Eq.~(\ref{eq2}) with $c_0\not=0$ to get the
 symmetry vector $X$. Thus, the task to obtain the conserved
 quantity $Q$ in Eq.~(\ref{eq6}) is also hard. Fortunately,
 there exists another way. Inspired by the discussion below
 Eq.~(\ref{eq82}), using Eqs.~(\ref{eq86}) and (\ref{eq87}),
 we find the conserved quantity as
 \be{eq88}
 \left(\mathring{\tilde{x}}^2-\frac{c_0}{n}
 \right)e^{2|n|\tilde{x}}=const.,
 \ee
 which can reduce to Eq.~(\ref{eq82}) if $c_0=0$. However,
 since we have no $f({\cal R})$ as an explicit function of
 $\cal R$ in this case, it is difficult to
 convert $\tilde{H}(\tilde{t})$ and $\tilde{a}(\tilde{t})$ into
 the cosmological solutions $H(t)$ and $a(t)$. Nevertheless, it
 is easy to see that $\tilde{H}(\tilde{t})$ and
 $\tilde{a}(\tilde{t})$ in Eqs.~(\ref{eq86}) and (\ref{eq87})
 are significantly different from the ones of power-law
 solution in Eqs.~(\ref{eq72}) and (\ref{eq73}). So, it is
 reasonable to speculate that $H(t)$ and $a(t)$ are also not
 power-law. In fact, we speculate that they might
 be hyperbolic, akin to $\tilde{H}(\tilde{t})$
 and $\tilde{a}(\tilde{t})$ in Eqs.~(\ref{eq86})
 and (\ref{eq87}). They are new solutions via Hojman symmetry.
 Anyway, let us turn to the case of $n=0$.


\subsubsection{The case of $n=0$}\label{sec3b2}

In the case of $n=0$, Eq.~(\ref{eq83}) becomes
 \be{eq89}
 2c_0 F^2 = F {\cal R}-2f\,,
 \ee
 which is still not easy to be solved directly by itself.
 However, we can try to indirectly solve it by the help of
 Eq.~(\ref{eq60}), whose left hand side is just the right hand
 side of Eq.~(\ref{eq89}). Note that
 $\rho_{_M}=\rho_{_{M0}}a^{-3}$ from Eq.~(\ref{eq12}) with
 $p_{_M}=0$. Using Eqs.~(\ref{eq89}) and (\ref{eq60}), we have
 $2c_0 F^2=-\kappa^2\rho_{_{M0}}a^{-3}$. So, $c_0<0$ is
 required. And then, we obtain
 \be{eq90}
 F=c_4\,a^{-3/2}\,,
 \ee
 where $c_4^2=-\kappa^2\rho_{_{M0}}/(2c_0)>0$ is constant.
 Substituting Eq.~(\ref{eq90}) into Eq.~(\ref{eq89}), we get
 \be{eq91}
 2c_0 c_4^2\,a^{-3}=c_4\,a^{-3/2}\,{\cal R}-2f\,.
 \ee
 Differentiating Eq.~(\ref{eq91}) with respect to $a$,
 and noting that $f_{,a}=f_{,\cal R}{\cal R}_{,a}=F{\cal R}_{,a}\,$,
 it is easy to obtain
 \be{eq92}
 2a^{5/2}\,{\cal R}_{,a}+2a^{3/2}\,{\cal R}=12c_0 c_4\,,
 \ee
 which is a differential equation for $\cal R$ with respect
 to $a$. Its solution reads
 \be{eq93}
 {\cal R}(a)=\left(6c_0 c_4\ln a+c_{10}\right)a^{-3/2}\,,
 \ee
 where $c_{10}$ is an integration constant. Substituting
 Eq.~(\ref{eq93}) into Eq.~(\ref{eq91}), we have
 \be{eq94}
 f(a)=\frac{1}{2}a^{-3}\left(6c_0 c_4^2\ln a
 +c_4 c_{10}-2c_0 c_4^2\right)\,.
 \ee
 Solving Eq.~(\ref{eq93}), we find
 \be{eq95}
 a^{-3/2}=-\frac{{\cal R}}{4c_0 c_4 W(c_1{\cal R})}\,,
 \ee
 where $c_1=-\exp(-c_{10}/(4c_0 c_4))/(4c_0 c_4)$ is constant,
 and $W(z)$ is the Lambert W function (or product
 logarithm)~\cite{r43}, which gives the principal solution for
 $w$ in $z=we^w$. Substituting Eq.~(\ref{eq95}) into
 Eq.~(\ref{eq91}), we obtain $f({\cal R})$ as an explicit
 function of $\cal R$, namely
 \be{eq96}
 f({\cal R})=-\frac{{\cal R}^2}{16c_0}\left[
 \,\frac{1}{W^2(c_1{\cal R})}+\frac{2}{W(c_1{\cal R})}\,\right] \,.
 \ee
 In fact, one can check that this $f({\cal R})$ indeed
 satisfies Eq.~(\ref{eq89}). {\em To our knowledge, this
 $f({\cal R})$ has not been considered previously
 in the literature.}

In the case of $c_0\not=0$ and $n=0$, substituting
 Eq.~(\ref{eq69}) into Eq.~(\ref{eq65}), we find
 that $\mathring{\tilde{H}}=c_0$, whose solution is given by
 \be{eq97}
 \tilde{H}(\tilde{t})=c_0\,\tilde{t}+c_2\,,
 \ee
 and then
 \be{eq98}
 \tilde{a}(\tilde{t})=c_3\exp\left(\frac{c_0}{2}
 \,\tilde{t}^{\,2}+c_2\tilde{t}\,\right)\,,
 \ee
 where $c_2$ and $c_3$ are both integration constants.
 Substituting Eq.~(\ref{eq90}) into $\tilde{a}=\sqrt{F}\,a$
 from Eq.~(\ref{eq13}), and using Eq.~(\ref{eq98}), it is
 easy to obtain
 \be{eq99}
 a=c_3^4 c_4^{-2}\exp\left(2c_0
 \tilde{t}^{\,2}+4c_2\tilde{t}\,\right)\,.
 \ee
 Substituting Eqs.~(\ref{eq90}) and (\ref{eq99})
 into $d\tilde{t}=\sqrt{F}\,dt$ from Eq.~(\ref{eq13}), we get
 \be{eq100}
 \frac{d\tilde{t}}{dt}=c_{51}\exp\left(\frac{3}{2}c_{00}
 \,\tilde{t}^{\,2}-3c_2\tilde{t}\,\right)\,,
 \ee
 where $c_{51}=\sqrt{c_4}\,c_3^{-3}c_4^{3/2}$ and
 $c_{00}=|c_0|=-c_0>0$ are both constants. From
 Eq.~(\ref{eq100}), we find that
 \be{eq101}
 \tilde{t}=\frac{c_2}{c_{00}}+\sqrt{\frac{2}{3c_{00}}}
 \,\Psi\left(c_5(t-t_0)\right)\,,
 \ee
 where $t_0$ is an integration
 constant, $c_5=c_{51}(6c_{00}/\pi)^{1/2}\exp(-3c_2^2/(2c_{00}))$ is
 also constant, and $\Psi(z)$ is the inverse error function
 ${\rm erf}^{-1}(z)$~\cite{r44}. Substituting Eq.~(\ref{eq101})
 into Eq.~(\ref{eq99}), we have
 \be{eq102}
 a(t)=c_6\exp\left(-\frac{4}{3}\Psi^2(c_5(t-t_0))\right)\,,
 \ee
 where $c_6=c_3^4 c_4^{-2}\exp(2c_2^2/c_{00})$ is constant.
 {\em Obviously, it is not a power-law solution. To our
 knowledge, this new solution has not been found previously in
 the literature.} From Eq.~(\ref{eq102}), it is easy to get the
 Hubble parameter as
 \be{eq103}
 H(t)=\frac{\dot{a}}{a}=-\frac{4}{3}c_5\sqrt{\pi}
 \,\Psi(c_5(t-t_0))\,\exp\left(\Psi^2(c_5(t-t_0))\right)\,.
 \ee
 From Eqs.~(\ref{eq102}) and (\ref{eq103}), we find that
 \be{eq104}
 H(a)=\pm\frac{2c_5\sqrt{\pi}\,a}{3c_6}\sqrt{-3\ln\frac{a}{c_6}}\,.
 \ee

On the other hand, it is hard to solve Eq.~(\ref{eq2}) with
 $c_0\not=0$ to get the symmetry vector $X$. Thus, the task to
 obtain the conserved quantity $Q$ in Eq.~(\ref{eq6}) is also
 hard. Fortunately, there exists another way. Inspired by the
 discussion below Eq.~(\ref{eq82}), using Eqs.~(\ref{eq97}) and
 (\ref{eq98}), we get the conserved quantity as
 \be{eq105}
 \mathring{\tilde{x}}^2-2c_0\tilde{x}=const.
 \ee


\section{Concluding remarks}\label{sec4}

Nowadays, $f(R)$ theory has been one of the leading modified
 gravity theories to explain the current accelerated expansion
 of the universe, without invoking dark energy. It is of
 interest to find the exact cosmological solutions of $f(R)$
 theories. Besides other methods, symmetry has been proved as
 a powerful tool to find exact solutions. On the other hand,
 symmetry might hint the deep physical structure of a theory,
 and hence considering symmetry is also well motivated. As is
 well known, Noether symmetry has been extensively used in
 physics. Recently, the so-called Hojman symmetry was also
 considered in the literature. Unlike Noether conservation
 theorem, the symmetry vectors and the corresponding conserved
 quantities in Hojman conservation theorem can be obtained by
 using the equations of motion directly, without using
 Lagrangian or Hamiltonian. In general, its
 conserved quantities and the exact solutions can be quite
 different from the ones using Noether symmetry.

In this work, we consider Hojman symmetry in $f(R)$ theories
 in both the metric and Palatini formalisms, and find the
 corresponding exact cosmological solutions of $f(R)$ theories
 via Hojman symmetry. The main difficulty to consider Hojman
 symmetry in the metric $f(R)$ theory is that the corresponding
 equations of motion are 4th order with respect to the scale
 factor $a$, while Hojman symmetry deals with 2nd order
 equations. We should try to recast them as 2nd
 order differential equations. Inspired by the well-known conformal
 transformation, we introduce new variables $\tilde{t}$ and
 $\tilde{a}$. This is the key idea to use Hojman symmetry
 in $f(R)$ theories. While the traditional conformal transformation
 mainly deals with the Lagrangian/action, here we instead
 directly deal with the equations of motion using this variable
 transformation. In the Palatini $f(R)$ theory, this transformation
 is also employed, although the corresponding equations of motion
 are already 2nd order with respect to the scale factor $a$.
 We find that the equations of motion can be significantly
 simplified by using this variable transformation.

In both the metric and Palatini $f(R)$ theories, we can obtain
 the power-law cosmological solutions via Hojman symmetry, as
 shown in Secs.~\ref{sec2a} and \ref{sec3a}. Note that such
 kind of power-law solutions can also be found by using Noether
 symmetry in both the metric and Palatini $f(R)$ theories (see
 e.g.~\cite{r21,r22,r23,r24,r25}). However, some new solutions
 significantly different from the power-law solution are also
 obtained by using Hojman symmetry, as shown in
 Secs.~\ref{sec2b} and \ref{sec3b}. {\em In fact, these new
 results cannot be found via Noether symmetry. To
 our knowledge, they also have not been found previously in
 the literature.}

Note that in the metric $f(R)$ theory, to be simple, we have
 only considered the ``dark energy'' dominated
 epoch following~\cite{r33,r34}, and hence
 the contributions from matter can be ignored. If we do not
 ignore the contributions from matter, from Eq.~(\ref{eq24}),
 it is easy to see that Eq.~(\ref{eq25}) should be changed to
 $\stackrel{\,_{\circ\circ}}{\tilde{x}}\ =-s(\tilde{x})\,
 \mathring{\tilde{x}}^2+\sigma(\tilde{x},\mathring{\tilde{x}})
 ={\cal F}(\tilde{x},\mathring{\tilde{x}})$, where the
 additional term $\sigma(\tilde{x},\mathring{\tilde{x}})$
 comes from $\tilde{\rho}_{_M}=\rho_{_M}/F^2$ and
 $\tilde{p}_{_M}=p_{_M}/F^2$. The condition~(\ref{eq5}), the
 equation for the symmetry vector $X$ (namely Eq.~(\ref{eq2}))
 and the equation used to derive $V(\phi)$ (namely Eq.~(\ref{eq29}))
 should become more complicated. It might be a tough work to
 solve these equations, and we leave it as an open question.

On the other hand, in the Palatini $f(R)$ theory, we do not
 ignore the contributions from matter. To be simple, in
 Sec.~\ref{sec3} we have only considered the case
 of $w_{_M}=p_{_M}/\rho_{_M}=0$, namely pressureless matter.
 In fact, one can consider a more general case of
 $w_{_M}=const.$ further. In Eqs.~(\ref{eq60})
 and (\ref{eq61}), one can incorporate $p_{_M}=w_{_M}\rho_{_M}$
 into the term $\rho_{_M}$, namely the right hand sides of
 Eqs.~(\ref{eq60}) and (\ref{eq61}) become
 $(1\pm 3w_{_M})\rho_{_M}$. Similarly, a factor $1+w_{_M}$
 appears in Eq.~(\ref{eq12}). And then, such kind of constant
 factors containing $w_{_M}$ can be absorbed by redefining the
 other constants in the model, as done in~\cite{r35}. So, it
 is reasonable to anticipate that the main results will not
 be affected, except some constants in the model might be
 rescaled, as shown in~\cite{r35}. The detailed calculations
 are straightforward and trivial, and hence we do not present
 them here.

Finally, we would like to briefly compare the cosmological
 results via Noether and Hojman symmetries in
 various scenarios. It is worth noting that the exact solutions
 of $f(R)$ theory via Noether symmetry have been discussed in
 the comprehensive review~\cite{r48} (we thank the referee for
 pointing out this issue). However, to our knowledge, the exact
 solutions of $f(R)$ theory in~\cite{r48} were
 mainly obtained in the spherically symmetric spacetime described by
 \be{eq106}
 ds^2=-A(r)\,dt^2+B(r)\,dr^2+M(r)\,d\Omega^2_2\,,
 \ee
 see e.g. Sec.~15.2 of~\cite{r48} (and~\cite{r49}) for details.
 Note that $M(r)=r^2$ and $A(r)=B^{-1}(r)=1-2{\cal M}/r$
 corresponds to the Schwarzschild case of GR. The general
 discussions on Noether symmetry in~\cite{r48} is very
 inspiring. However, the exact cosmological solutions in the
 FRW spacetime described by Eq.~(\ref{eq8}) have not been
 explicitly discussed in~\cite{r48}. Nevertheless, the works
 in~\cite{r48,r49} inspire us to further consider the exact
 solutions in the spacetime described by Eq.~(\ref{eq106})
 via Hojman symmetry instead, and we leave it to the future
 works. On the other hand, the same authors of~\cite{r48} have
 indeed considered the exact cosmological solutions of the
 metric $f(R)$ theory via Noether symmetry in the FRW spacetime
 described by Eq.~(\ref{eq8}) in a series of works (see
 e.g.~\cite{r50,r22}). They found that both the
 exact cosmological solutions and the functional form of $f(R)$ are
 power-law (see also~\cite{r21}). In e.g.~\cite{r24,r25}, the exact
 cosmological solutions of the Palatini $f(R)$ theory via
 Noether symmetry in the FRW spacetime have also been found. Again,
 these solutions and the functional form of $f(R)$ are
 also power-law. However, in the present work, we find that
 the exact cosmological solutions and the functional form of
 $f(R)$ can be {\em not} power-law, by using Hojman symmetry
 in both the metric and Palatini formalisms, as shown in
 Secs.~\ref{sec2b} and \ref{sec3b}.
 In addition, as mentioned in Sec.~\ref{sec1}, it is found that
 Hojman symmetry exists for a wide range of the potential
 $V(\phi)$ of quintessence~\cite{r33} and scalar-tensor
 theory~\cite{r34}, and the corresponding exact cosmological
 solutions have been obtained. While Noether symmetry exists
 only for exponential potential $V(\phi)$~\cite{r19,r26,r27},
 Hojman symmetry can exist for a wide range of potentials
 $V(\phi)$, including not only exponential but also power-law,
 hyperbolic, logarithmic and other complicated
 potentials~\cite{r33,r34}. On the other hand, it is also
 found that Hojman symmetry exists in $f(T)$ theory and the
 corresponding exact cosmological solutions are
 obtained~\cite{r35}. The functional form of $f(T)$ is
 restricted to be the power-law or hypergeometric type, while
 the universe experiences a power-law or hyperbolic expansion.
 These results are also different from the ones obtained
 by using Noether symmetry in $f(T)$ theory~\cite{r28}.
 So, in summary, Hojman symmetry can bring new features to
 cosmology and gravity theories.


\section*{ACKNOWLEDGEMENTS}
We thank the anonymous referee for quite useful comments and
 suggestions, which helped us to improve this work. We are
 grateful to Profs. Rong-Gen~Cai and Shuang~Nan~Zhang for
 helpful discussions. We also thank Minzi~Feng, as well as
 Shoulong~Li, Ya-Nan~Zhou and Dong-Ze~Xue for kind help
 and discussions. This work was supported in part by NSFC
 under Grants No.~11575022 and No.~11175016.

\renewcommand{\baselinestretch}{1.0}


\end{document}